\documentclass[useAMS,usenatbib]{mn2e}
\usepackage{graphicx}
\usepackage{natbib}

\title[MW disc response to LMC]{Response of the Milky Way's disc to the Large Magellanic Cloud in a first infall scenario}
\author[Laporte, Gomez, Besla, Johnston \& Garavito]{
\parbox[t]{\textwidth}{Chervin F. P. Laporte$^{1}$, Facundo A. G\'omez$^{2}$, Gurtina Besla$^{3}$, Kathryn V. Johnston$^{1}$, Nicolas Garavito-Camargo$^{3}$}
\\
\\
$^{1}$ Department of Astronomy, Columbia University, 550 West 120th Street, New York, NY, 10027, U.S.A\\
$^{2}$ Max-Planck-Institut f{\"u}er Astrophysik, Karl-Schwarzschild-Str. 1, D-85748, Garching bei M{\"u}enchen, Germany\\
$^{3}$ Steward Observatory, University of Arizona, 933 North Cherry Avenue, Tucson, AZ, 8572, U.S.A.\\
}
\begin{document}
\date{}

\pagerange{\pageref{firstpage}--\pageref{lastpage}} \pubyear{2011}
\maketitle
\label{firstpage}
\begin{abstract}

We present N-body and hydrodynamical simulations of the response of the Milky Way's baryonic disc to the presence of the Large Magellanic Cloud during a first infall scenario. For a fiducial galactic model reproducing the gross properties of the Galaxy, we explore a set of six initial conditions for the LMC of varying mass which all evolve to fit the measured constraints on its current position and velocity with respect to the Galactic Center. We find that the LMC can produce strong disturbances - warping of the stellar and gaseous discs - in the Galaxy, without violating constraints from the phase-space distribution of stars in the Solar Neighbourhood. All models correctly reproduce the phases of the warp and its anti-symmetrical shape about the disc's mid-plane. If the warp is due to the LMC alone, then the largest mass model is favoured ($2.5\times10^{11}\,\rm{M_{\odot}}$). Still, some quantitative discrepancies remain, including deficits in height of $\Delta Z=0.7  \,\rm{kpc}$ at $R=22 \,\rm{kpc}$ and $\Delta Z=0.7\, \rm{kpc}$ at $R=16 \,\rm{kpc}$.  This suggests that even higher infall masses for the LMC's halo are allowed by the data. A comparison with the vertical perturbations induced by a heavy Sagittarius dSph model ($10^{11}\,\rm{M_{\odot}}$) suggest that positive interference with the LMC warp is expected at $R=16 \, \rm{kpc}$. We conclude that the vertical structure of the Galactic disc beyond the Solar Neighbourhood may jointly be shaped by its most massive satellites. As such, the current structure of the MW suggests we are seeing the process of disc heating by satellite interactions in action.


\end{abstract}
\begin{keywords}
galaxies: formation - galaxies: evolution 
\end{keywords}

\section{Introduction}
Observations of HI in galaxies reveal that their outer gaseous discs are often warped, with their structures being typically anti-symmetrical, but varying significantly in height from one galaxy to another (e.g. \cite{binney92} for a review and references therein, see also \citet{reshetnikov16}). Multiple explanations have been proposed for the existence of warps. These have ranged from misaligned halos with respect to the disc, cosmic infall, misaligned accretion of cold gas (out of the disc plane) and tidal interactions with satellites \citep{debattista99,jiang99,roskar10, weinberg98, aumer13, gomez15b, gomez16}. Although warps were originally thought to be long-lived normal modes of the disc \citep{sparke88}, later analytical studies considering the effect of dynamical friction between the warp and the halo showed that warps are damped/excited by the dark matter halo on timescales shorter than a Hubble time, indicating the warps must be recently or continuously excited \citep{nelson95}. This is in line with simulations of \cite{shen06} who showed that warps excited by interactions with satellites can live for a few Gyrs, after which other infall events may occur, exciting new features. 

In the case of the Milky Way (MW), there are two known satellites that are massive and close enough to the disc - Sagittarius dSph ($20 \, \rm{kpc}$; Sgr) and the Large Magellanic Cloud ($50 \,\rm{kpc}$; LMC) - such that their gravitational field may significantly perturb the MW's dark halo and disc (e.g. \citet{weinberg98,jiang99,bailin03, purcell11, gomez13}). Likewise, M31 also shows signs of interactions with some of its satellites, such as M32 \citep{dierickx14}.

The warp of the MW's disc has been observed in neutral HI gas, dust and stars \citep{henderson82,levine06, drimmel01, djorgovski89,momany06} but each tracer is sensitive to very different radial extents. At a distance of $R\sim20 \,\rm{kpc}$,the HI warp is found to be asymmetric, reaching a height $\sim 4 \,\rm{kpc}$ above the midplane in the North ($l\sim 90$) and curving to the South ($l\sim270$) below $1\, \rm{kpc}$ \citep{levine06}. Although the mass of the neutral gas is much smaller than the mass of the stellar disc, it is much more extended and its distribution has been mapped well beyond the solar neighbourhood, out to $\sim30 \, \rm{kpc}$, with extensive angular coverage. On the other hand the dust maps from COBE/DIRBE, as well as observations of individual stars, have been primarily confined to regions close to the solar neighbourhood \citep{drimmel01,momany06, reyle09} and have typically considered S-shaped warps in the modeling. These studies find, differences in the amplitudes of the warp for all three tracers, increasing in the following order: HI, dust, stars (see \citet{reyle09}). These differences have been suggested to be evidence of the importance of hydrodynamical and magnetic forces \citep{drimmel01}. For example, \cite{reyle09} found that a parametrising the stellar disc using the solution for the warped HI disc of Levine et al. 2006b, did not produce an accurate description of the vertical structure of the disc.

The nature of this discrepancy may be related to the complex structure of the stellar disc, which is not only warped, but also exhibits structures such as rings \citep{newberg03,morganson16} or ripples \citep{xu15,price-whelan15}. In the solar neighbourhood, these structures have been observed in SDSS data through vertical asymmetries in number density counts and stellar radial velocities \citep{widrow12}. The RAVE and LAMOST surveys have also confirmed the asymmetries in velocity space \citep{williams13,carlin13}, attributing the features as signatures of a breathing mode. The existence of such perturbations may seriously impact our inferences of the local dark matter density \citep{banik16}. While both interactions with satellites \citep{widrow14} or non-axisymmetric features in the disc, such as the bar or spiral arms \citep{debattista14,siebert14, monari15,monari16} can give rise to breathing modes in the disc, the fact that bending modes are also observed in the number counts favours an interaction origin. These vertical spatial asymmetries persist on larger scales beyond the solar radius and amplify with increasing distance such that they extend well above the midplane, allowing them to be detected as overdensities in optical stellar surveys \citep{newberg03, ibata03, martin07, slater14, xu15, morganson16, bernard16}. These include the Monoceros ring \citep{newberg03, morganson16} and possibly the Triand overdensities \citep{ibata03,martin07, xu15} which extend to heights of $Z\sim-10\,\rm{kpc}$ out to $R\sim30\, \rm{kpc}$, all of which may be part of the disc. In the latter case, \citet{price-whelan15} brings further credence to this by looking at the stellar populations associated with the Triand I \& II clouds (in particular the ratio between red giants and RR Lyrae) finding that they are consistent with belonging to the disc. Throughout this paper we will term all vertical disturbances in physical and velocity space on smaller scales than the warp in the disk by the generic term ``ripples''.

In the tidal interaction scenario, warps result from the response of a galactic disc to the normal modes produced by the wake formed in the host dark matter halo and the orbiting satellite itself \citep{weinberg98}. Early N-body models found contradicting results in the amplitude as well as the location for the lines of nodes of the warp \citep{garcia-ruiz02,tsuchiya02}. The discrepancy arose from particle noise in low resolution N-body simulations. \cite{weinberg07} found that of order millions of particles are required to capture the resonant dynamics between the satellite and its host halo in producing the wake necessary to torque a disc. Using high-resolution fully cosmological simulations \cite{gomez15b} showed that indeed this mechanism alone can induce very strong vertical perturbations on a pre-existing disc.

In the most up-to-date study on the warp of the MW, \cite{weinberg06} calculated the disc response to the passage of the LMC traveling on a rosette orbit, with multiple pericentric approaches, using the perturbative method developed in \cite{weinberg98}. They found general qualitative agreement with the observational data of the MW's HI disc \citep{levine06}. 

However, a re-assessment is motivated due to the recent proper motion measurements of the Clouds \citep{kallivayalil06, kallivayalil13} suggesting that the LMC is on a first infall orbit \citep{besla07}. This could imply that the LMC is much more massive than the $1-3 \times 10^{10} \, \rm{M_{\odot}}$ typically assumed in the literature \citep{kerr57,burke57,hunter69,weinberg06}. Indeed, different lines of evidence favour a higher mass. Within the current cosmologically favoured model $\Lambda$CDM, a pair as massive as the LMC/SMC is most likely to be found on a first infall in MW-like host dark matter halos \citep{boylan-kolchin11}. 
The longevity of bound, binary LMC/SMC pair is also facilitated by a higher LMC mass \citep{besla12}. Finally, dynamical timing arguments also indicate that the LMC may be more massive than previously assumed \citep{penarrubia16}. These independent lines of evidence call for a re-assessment of the role played by the LMC in producing normal bending modes within the disc and setting the present structure of the neutral hydrogen gas as well as the stars within our MW. 

Moreover, the Sagittarius dSph is moving in a plane perpendicular to that of the LMC and has been around for a longer time. Though less massive, it has smaller pericenter. This raises the question as to whether the effect of Sagittarius alone or the combination of the two satellites could help reproduce all the complex features seen in both the optical and gaseous disc\footnote{The Small Magellanic Cloud (SMC) is traveling on orbit close to the same as that of the LMC and would excite the same resonances. Its inclusion would be somewhat equivalent to studying a model with an increased LMC mass.}. Recent advances in computing power have also made this problem tractable with the use of multi-million particle N-body simulations \citep{aumer13,yurin14,gomez15b}. 

In this study we use high resolution N-body simualtions to quantify the gravitational impact of the LMC on the galactic disc through tidal interactions and the wake produced on the MW's dark matter halo on a first infall. We explore a larger range of LMC masses than traditionally explored in previous studies, with the goal of constraining this mass range by examining LMC models that produce unrealistic responses in the MW disc. In section 1, we present the numerical methods employed in setting our initial conditions for our experiment. In section 2 we present results from purely collisionless N-body runs and present results with the inclusion of a cold gaseous component. We also contrast our results with runs considering the effect of the Sagittarius galaxy with an initial mass of $m_{Sgr}=1\times10^{11} {\rm{M_{\odot}}}$ in section 3. We finish with a discussion and conclude in section 4.

\section{Numerical methods}

The N-body simulations presented here were carried out with the {\sc gadget-3} code \citep{Springel2005a}. In this section we describe the mass models used for the LMC and the MW as well as the orbits on which each realisation was initialised with.

\subsection{Models of the LMC}

We consider 6 models for the LMC of varying virial masses $M_{200}$\footnote{defined as the mass enclosed within a radius $r_{200}$ containing 200 times the critical density $\rho_{c}$ of the Universe} in the range $3.0\times 10^{10} \rm{M_{\odot}} - 2.5 \times 10^{11} {\rm{M_\odot}}$. The LMC halo is parametrised as \cite{Hernquist1990} profiles with masses $M$ and scale radii $a$ chosen such as to match the measured total mass at the optical radius $M(r<9)\sim1.7\times10^{10}{\rm M_{\odot}}$ \citep{vdmarel14} and the considered virial masses. These values are listed in Table 1. The Hernquist density profile can broadly match the commonly found NFW profile in cosmological simulations \citep{Navarro1996}, the only difference being that it has a steeper fall-off at large radii \citep{Jang2001}. The circular velocity curves and cumulative mass profiles for each model are shown in Figure 1. Note that the peak of their circular velocity vary from $90 \, \rm{km/s}$ and $110 \rm{km/s}$, which is within the observed range \citep{vdmarel02,vdmarel14}. The N-body models are generated by Monte-Carlo sampling the respective self-gravitating Hernquist distribution functions through standard methods \citep{Kuijken1994}. We do not model the contribution of the LMC's disc as we take the dynamical mass constraints into account in our halo parametrisation. 


\begin{table}
 \centering
 \begin{minipage}{130mm}
  \begin{tabular}{@{}llrrrrlrlr@{}}
  \hline
Run & $ M_{200}$ & M & a \\
 &   /$10^{10}\rm{M_{\odot}}$ & $/10^{10} \rm{M_{\odot}}$ & kpc \\
  \hline
1 & $3.00 $  & $3.4 $ & $ 3.3  $ \\
2 & $5.00 $  & $6.3 $ & $ 8.3  $ \\
3 & $8.00 $  & $10.7$ & $ 13.5 $ \\
4 & $10.0 $  & $13.9 $ & $ 16.8 $ \\
5 & $18.0 $  & $27.6$ & $ 27.2 $ \\
6 & $25.0 $  & $40.0$ & $ 34.6 $ \\
\hline
\end{tabular}
\end{minipage}
\caption[LMC ICs]{LMC structural properties for 6 models listing: virial masses $M_{200}$, total mass $M$ and scale radius $a$ used in the Hernquist parametrisation tailored to reproduce the listed virial masses and total mass within the optical radius $M(<9.0)\sim1.7\times10^{10} \, \rm{M_{\odot}}$.}
\end{table}

\begin{figure}
\includegraphics[width=0.5\textwidth,trim=0mm 0mm 0mm 0mm,clip]{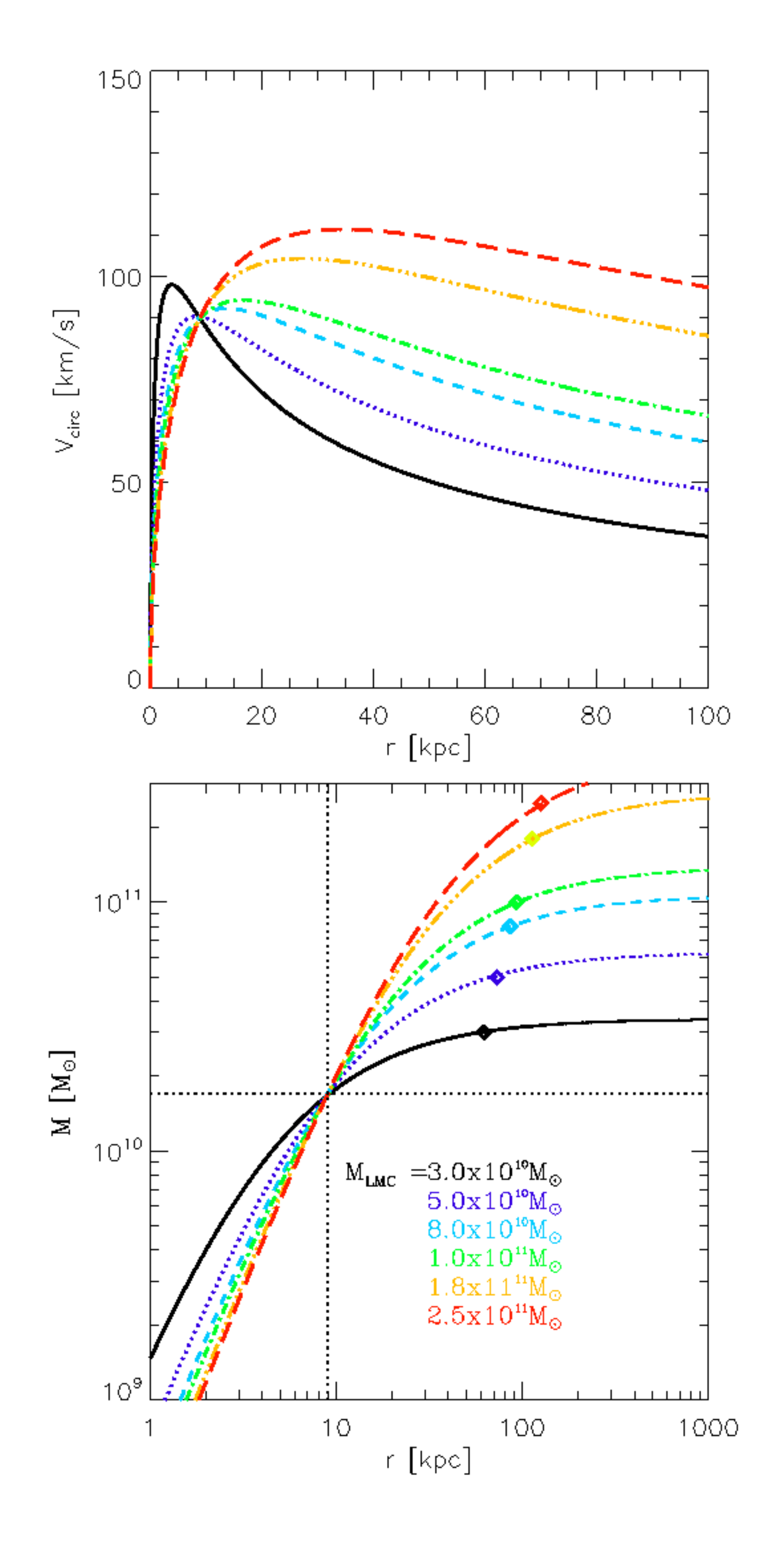}
\caption[]{{\it Top Panel}: Circular velocity curves as a function of radius  for the six LMC models considered in this study. The curves peak between $90\, {\rm km/s}$ and $110 \, \rm{km/s}$. {\it Bottom Panel:} Cumulative mass distribution as a function of radius. The mass within the optical radius is $M(<9.0)\sim1.7\times10^{10}\,\rm{M_{\odot}}$, which is common to all models. The diamonds mark the masses at the virial radius for each model, corresponding to $M_{200}$.}
\end{figure}


\begin{figure}
\includegraphics[width=0.5\textwidth,trim=0mm 0mm 0mm 0mm,clip]{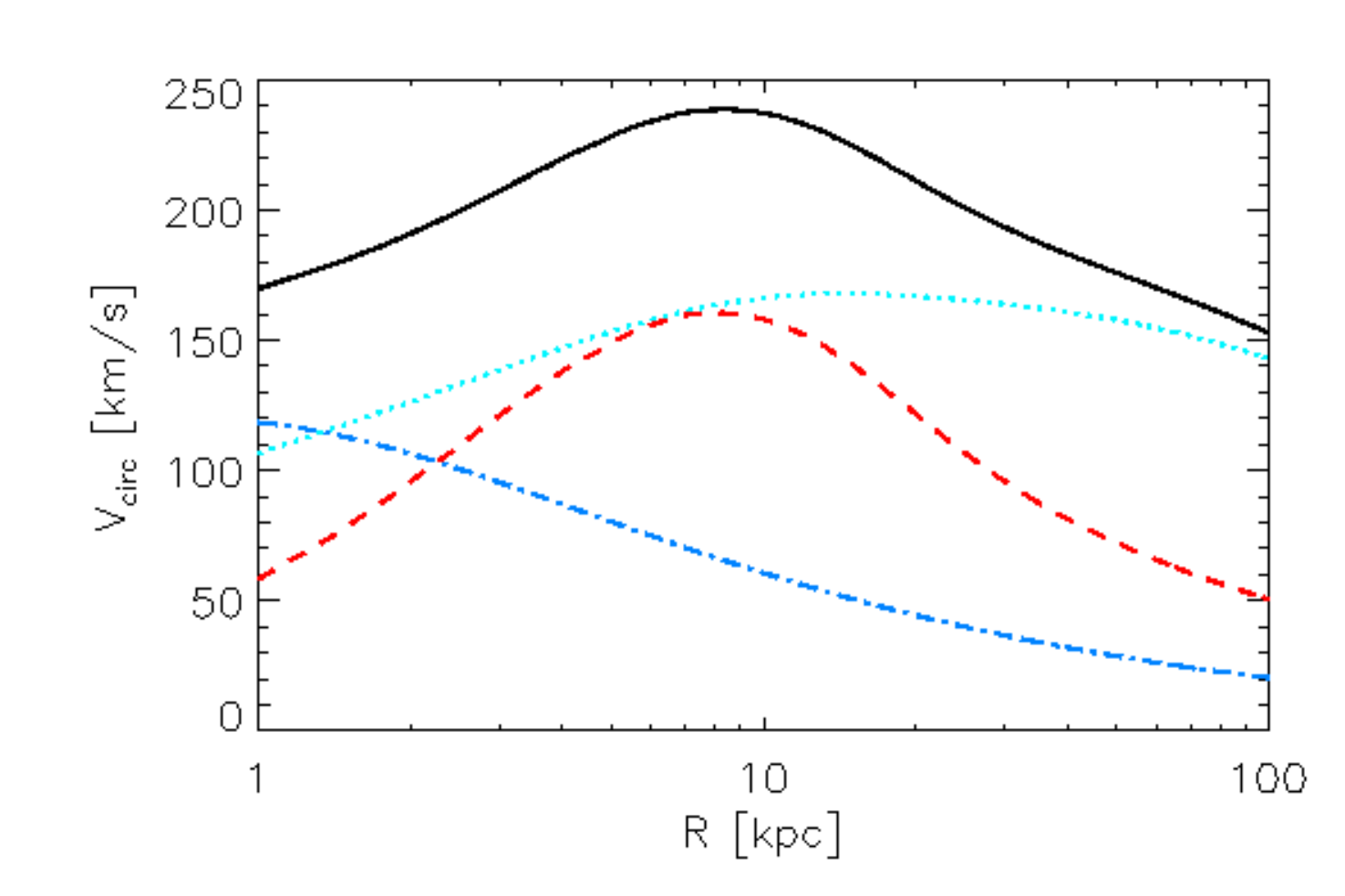}
\caption[]{Circular velocity curve for our MW model considered here (solid black line). This model has a circular velocity of 239 km/s at the solar radius $R\sim8\,\rm{kpc}$. The blue dash-dotted, red dashed and turquoise dotted lines represent the contributions of the bulge, disc and halo respectively.}
\end{figure}

\begin{table}
 \centering
 \begin{minipage}{130mm}
  \begin{tabular}{@{}llrrrrlrlr@{}}
  \hline
Components & & units \\
  \hline
  \hline
DM halo & & $N_{part}= 20,000,000$ \\
Virial mass &$1\times 10^{12}$ & $\rm{M_{\odot}}$ \\
Scale radius & $28$ & \rm{kpc}\\
Concentration & $10$ &\\
\hline 
Stellar disc & & $N_{part}=6,000,000$\\
Mass & $6.5 \times 10^{10}$&  $\rm{M_{\odot}}$\\
Scale length & $3.5$ & $\rm{kpc}$\\
Scale height & $0.53$& $\rm{kpc}$\\
\hline
Bulge && $N_{part}=1,000,000$\\
Mass & $1\times 10^{10}$ & $\rm{M_{\odot}}$\\
Scale radius & $0.7$ & $\rm{kpc}$\\
\hline
Gas disc & & $N_{part}=1,000,000$\\
Mass  & $8.7\times 10^{9}$ & $\rm{M_{\odot}}$\\
Scale length & $3.5$ & $\rm{kpc}$\\

\end{tabular}
\end{minipage}
\caption[LMC ICs]{Table summarizing the parameters used for our mass model of the MW. For our model, the circular velocity peaks at 239 km/s at the solar radius ($R=8.0\,\rm{kpc}$). }
\end{table}

\subsection{Model for the MW}
For the MW, we adopt a model close to the one used in \cite{gomez15} using the initial conditions generator {\sc galic} by \cite{yurin14}. The virial mass of the galaxy is $\sim10^{12} \rm{M_{\odot}}$ with a dark halo represented by a Hernquist sphere as in \citep{Springel2005c} with mass $M_{h}=9.3\times10^{11}\,\rm{M_{\odot}}$ and scale length $a_{h}=28\, \rm{kpc}$. The disc is modeled as an exponential disc with scale radius $R_{d}=3.5 \, \rm{kpc}$ and scale height $h_{d}=0.53 \,\rm{kpc}$ and a total mass $M_{d}=6.5\times 10^{10} \rm{M_{\odot}}$. The bulge is represented by a Hernquist sphere with mass $M_{b}=1\times10^{10} \rm{M_{\odot}}$ and a scale radius of $a=0.7 \,\rm{kpc}$. For these choices of parameters the rotation curve peaks at a value of $\sim 239 \rm{km/s}$ as parametrised by \cite{mcmillan11}. The circular velocity curve for our MW model is shown in Figure 2, where we also show the contribution of each separate component (dark halo, disc and bulge). When adding gas, the HI disc follows an exponential profile, scale length $R_{0}=3.5 \,\rm{kpc}$ and mass $M_{HI}=8.7\times 10^{9} \rm{M_{\odot}}$. The gas is set such that it feels the same total potential from the stars, bulge and dark halo just as in \citep{Springel2005c}. We note that the Toomre Q parameter of the disc is above unity everywhere except in a small region around $3-8 \,\rm{kpc}$ where it falls to $0.9$ at its lowest value. Although this is synonymous with the disc being susceptible to bar instability, our stability tests which were run for the same duration as in the presented N-body experiments ($t\sim2.0-2.1\, \rm{Gyr}$) did not go unstable. A bar forms only after $t=3.5\,\rm{Gyr}$ of evolution. The disc velocity structure is set such that the vertical streaming velocity is $\overline{v_{z}}=0$.



\subsection{Orbital parameters}

The orbits of the LMC-MW pair were re-calculated from \cite{gomez15} in order to take into account the effect of the LMC on the MW's centre of mass's motion. Briefly, the orbits of both the LMC and MW are evaluated through backwards time integration from the present-day positions and velocities with an implementation of dynamical friction for a body moving through an infinite homogeneous medium \citep{chandrasekhar43}. The semi-analytic orbit calculation is then re-calibrated with the aid of low resolution simulations. The infall time for the LMC is at about $t=2$ Gyr in the past. We use the position and velocity of each galaxy with respect to each other at that time to set the simulations for all models and run them to the present-day. Figure 3 shows the different orbits for the six LMC models considered here. The LMC orbit is on a first infall, primarily confined to the YZ plane with a pericentric approach occurring at 0.1 Gyr from the present-time. Currently it is just past pericenter $R\sim 50\, \rm{kpc}$ and is moving away from the MW.

\begin{figure}
\includegraphics[width=0.5\textwidth,trim=0mm 0mm 0mm 0mm,clip]{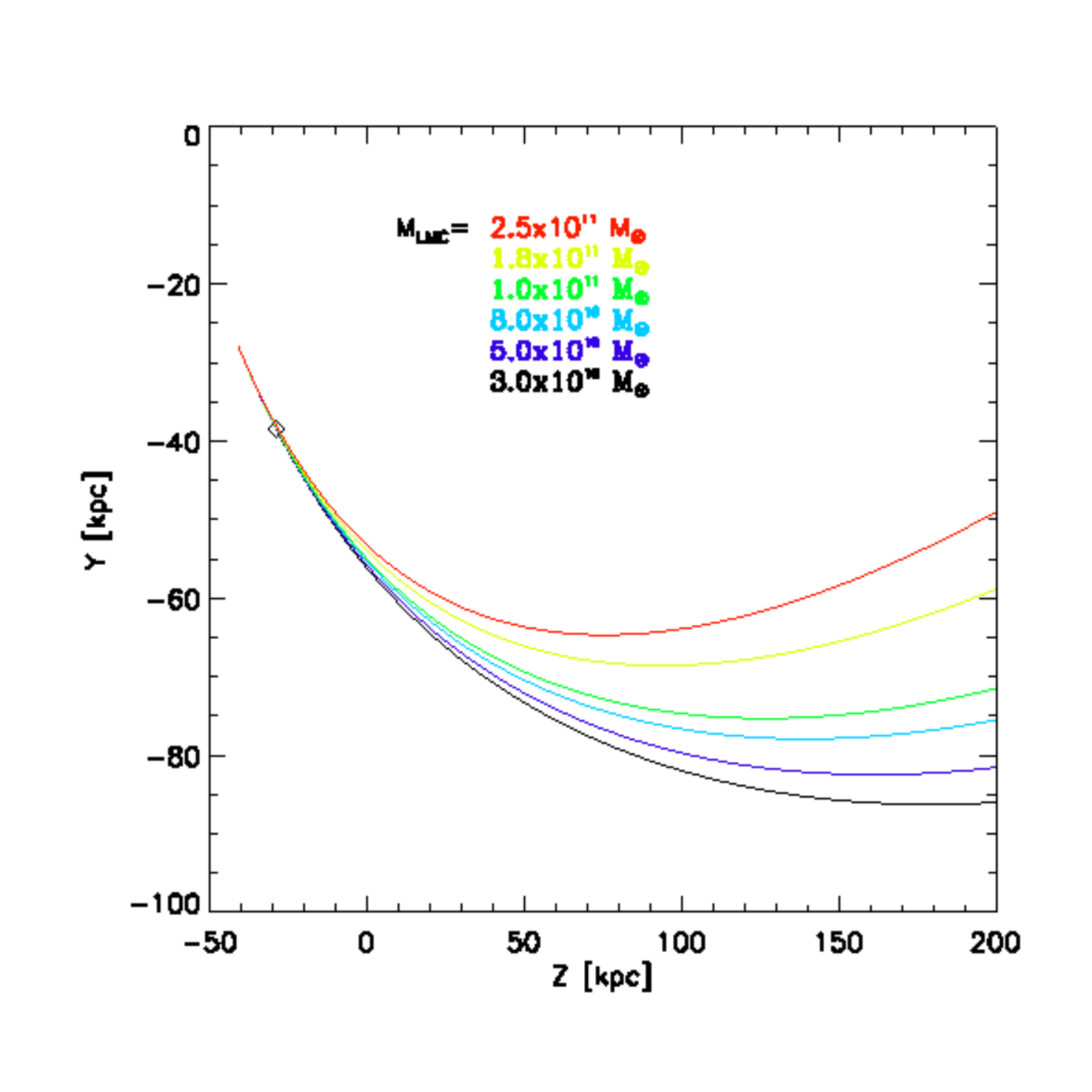}
\caption[]{Orbits for the six different LMC models considered. The black and red lines correspond to the least and most massive models respectively. In our coordinate system, the sun is located at (-8,0,0). The diamond represents the point of closest approach.}
\end{figure}

\section{Results}

In order to confirm that the orientation of the galactic disc relative to the LMC's orbital plane is correct in our models we compute the center of mass of the MW by using a shrinking sphere algorithm. We then align the galaxy onto the plane perpendicular to the disc's angular momentum vector by selecting all disk particles within a cylinder radius of $8 \, \rm{kpc}$ and rotating the system such that the $L_{z}$ component of the angular momentum is aligned with the $z$-axis. We decrease the height of the cylinder and repeat the procedure until convergence is met. Finally, the position and velocity of the LMC with respect to the MW is also determined through a separate shrinking sphere algorithm on its individual constituent particles. In particular, we make sure that the models fall systematically within $2\sigma$ from the uncertainty in distance ($\sigma_{D}=2.5 \,\rm{kpc}$) from \cite{freedman01} and speed ($\sigma_{v}=24 \rm{km/s}$) from \cite{kallivayalil13}. This is summarised in Table 2, where we list the uncertainties in all phase-space coordinates as well as in distance and speed of the LMC with respect to the Galactic center. Given the reasonable success of these models at reproducing both the present-day galactocentric distance and speed of the LMC, we now turn our attention to the tidal effect of the LMC on the disc of the MW.


\begin{table}
 \centering
 \begin{minipage}{130mm}
  \begin{tabular}{@{}llrrrrlrlr@{}}
  \hline
\# & $\Delta x$ & $\Delta y$ & $\Delta z$ & $\Delta v_{x}$ & $\Delta v_{y}$ & $\Delta v_{z}$ & $\Delta D$ & $\Delta v$\\
  \hline
1  &  -0.1  &  4.0  &  0.3 &   -11.0 &  -12.5  &   34.3  &   -3.5   &   34.7\\
2  &  -0.4  &  3.8  &  2.0 &     2.5  &    6.9    &   40.9  &   -4.3   &   24.5\\
3  &  -0.6  &  3.1  &  2.3 &     3.2  &    5.6    &   42.9  &   -3.8   &   26.8\\
4  &  -0.2  &  3.5  &  1.8 &    -1.8  &    7.7    &   42.8  &   -3.9   &   26.1\\
5  &  -1.6  & -0.6  &  5.8 &    -4.2  &   -16.5  &   50.4  &   -2.4   &   47.8\\
6  &  -1.1  & 0.0  &  3.9 &     3.9  &   -32.9  &   46.9  &   -2.0   &   55.0\\
\hline
\end{tabular}
\end{minipage}
\caption[]{Differences $\Delta=X_{sim}-X_{data}$ in position, velocity, position and speed between the model realisations and LMC data from \citep{kallivayalil13}. The adopted phase-space location of the LMC is taken to be $X_{data}=(-1.06, -41.0, -27.0, -57.4,-225.5, 220)$. The final distance and speeds for the various LMC models are within $2 \sigma$ from those determined observationally - $\sigma_{v}=24 \rm{km/s}$ \citep{kallivayalil13} and $\sigma_{D}=2.5\, \rm{kpc}$ \citep{freedman01} - except for the last model which slightly exceeds $2\sigma$.}
\end{table}

\begin{figure}
\includegraphics[width=0.5\textwidth,trim=0mm 0mm 0mm 0mm,clip]{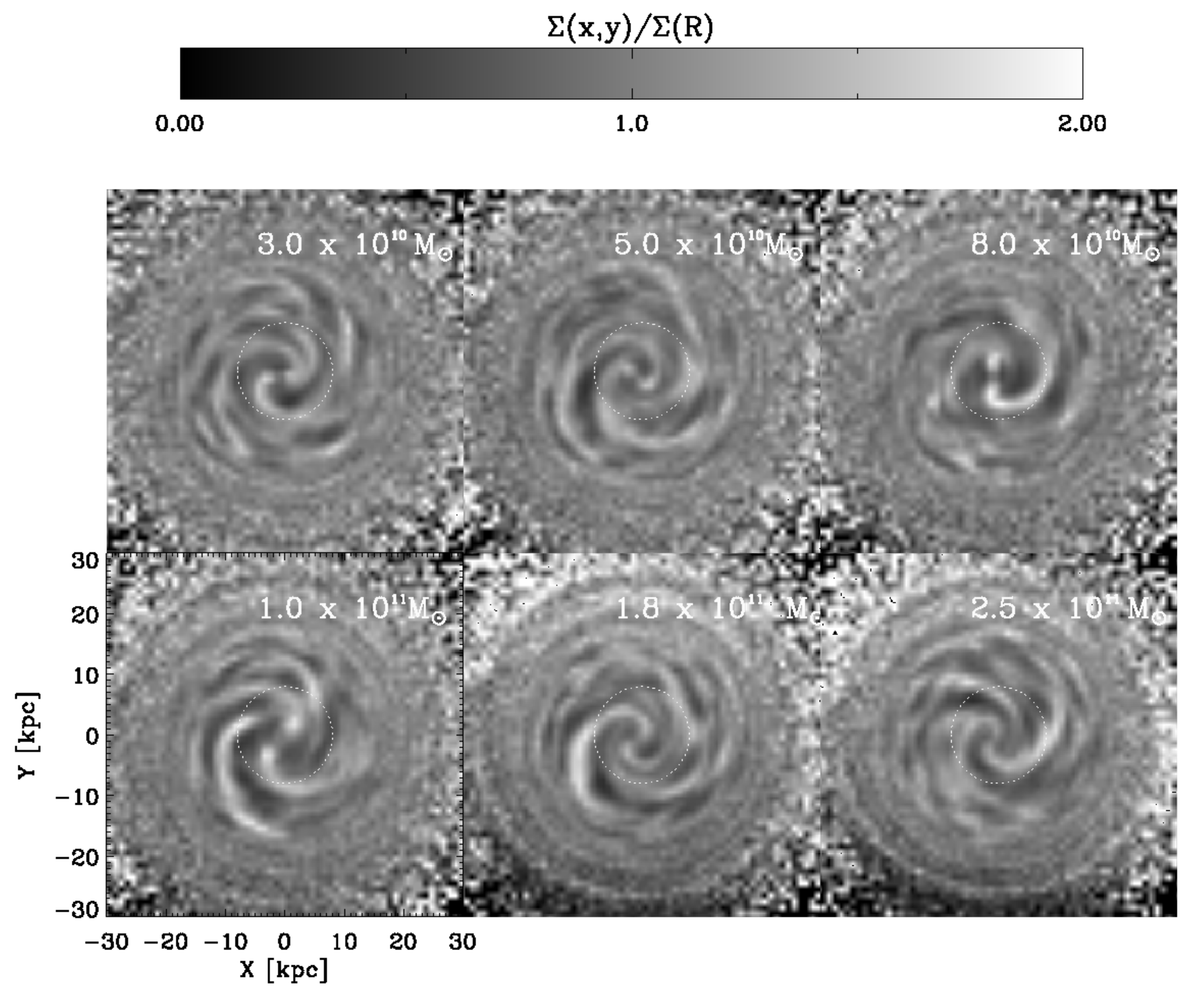}
\caption[]{Overdensity maps of the stellar disc at the present day for all six LMC models. Overall the disc is preserved through the interactions with mild spiral structure and a bar in some cases.}
\end{figure}

\begin{figure}
\includegraphics[width=0.5\textwidth,trim=0mm 0mm 0mm 0mm,clip]{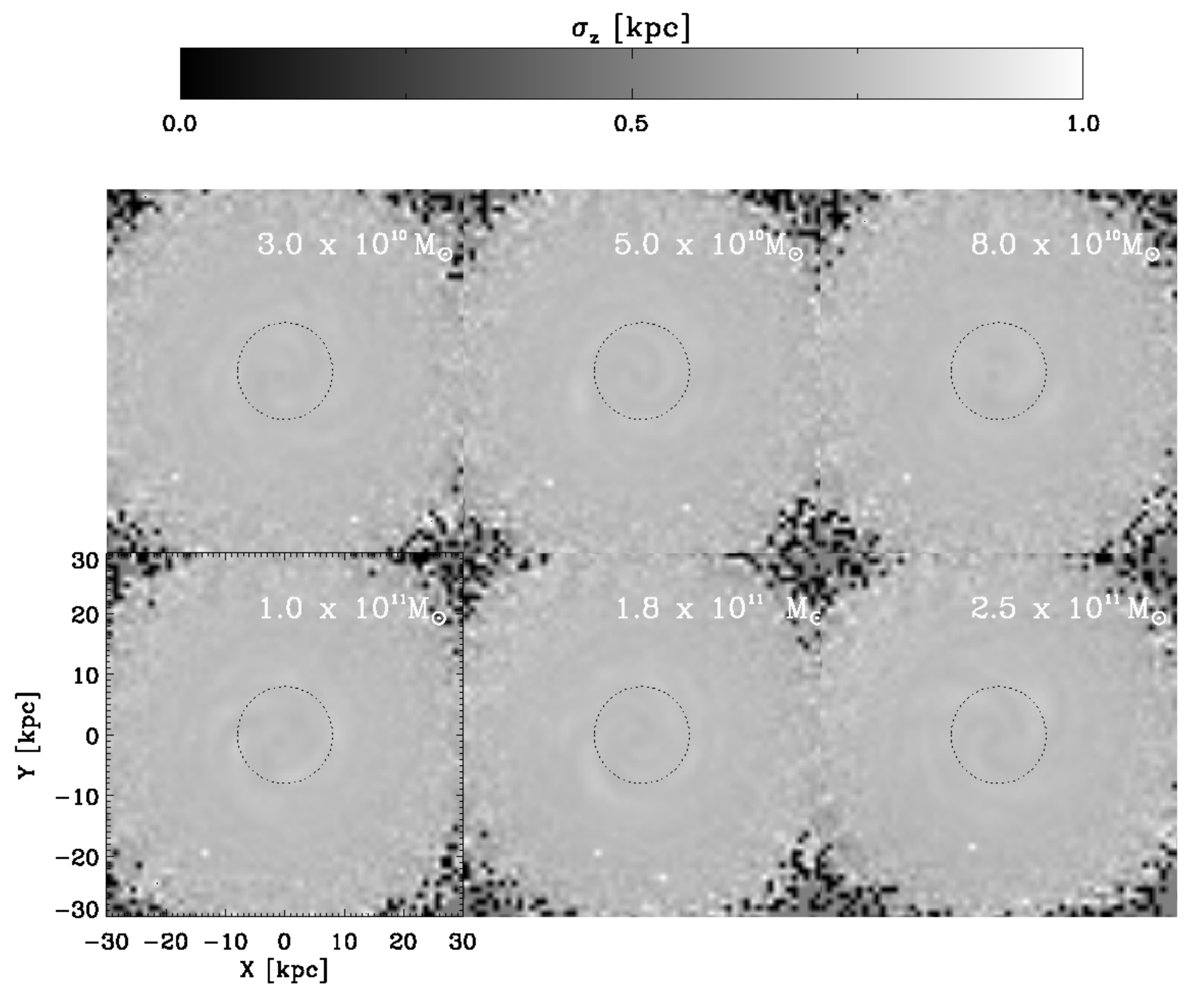}
\caption[]{Standard deviation of disc particle heights for all six simulations at the present-day.}
\end{figure}

\begin{figure}
\includegraphics[width=0.5\textwidth,trim=0mm 0mm 0mm 0mm,clip]{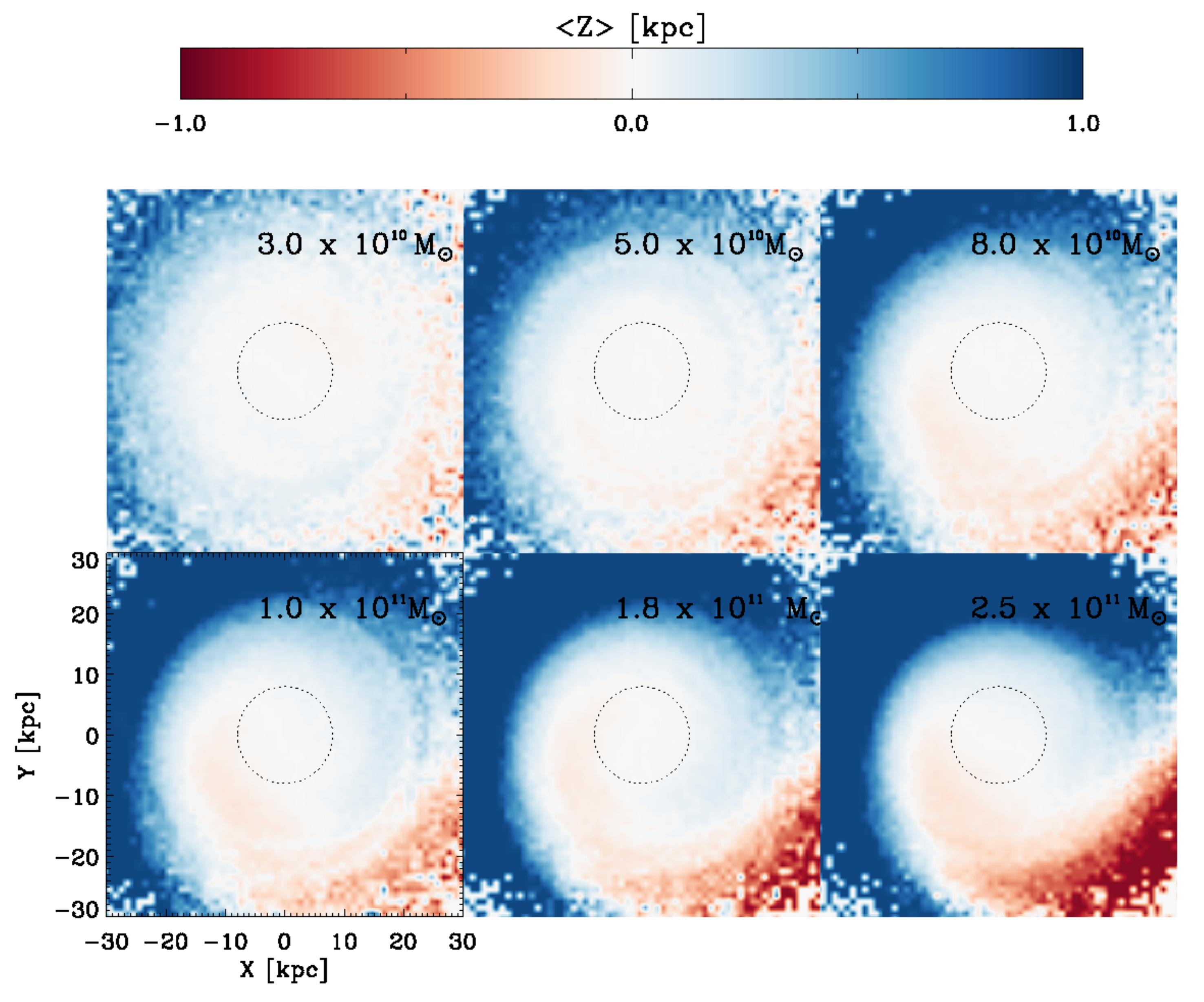}
\caption[]{Mean disc height maps for all six realisations. The solar circle at $R\sim8.0\,\rm{kpc}$ is denoted by the dashed circle. For the more massive models a warp in the stars can already be discerned within the solar neighbourhood.}
\end{figure}

\subsection{Response of the disc}

Before studying the warp of the simulated MW, it is important to check whether any LMC mass model can be readily ruled out (e.g. whether the disc is destroyed, excessively heated or a strong bar is formed that extends to the solar radius). By looking at the overdensity maps for all six collisionless realisations in Figure 4, we see that the disc is preserved in all interactions. Some mild spiral structures are discernable as well as bars in some cases. Figure 5 shows the vertical dispersion in stellar heights which is uniform across the disc showing that the disc was not overly heated/flared. \\

Figure 6 shows maps of the mean height of stars for each simulation which is computed by taking the mean of all the $z$-heights of the disc particles which fall within each pixel of $1 \,\rm{kpc}$ in size. Even the least massive model ($M_{LMC}=3\times10^{10}\,\rm{M_{\odot}}$) produces some warping with similar azimuthal phase but the amplitude of the warp varies as a function of satellite mass considered for the LMC. We also mark for clarity a circle at $R=8 \,\rm{kpc}$ to denote the solar radius. We note that material can be launched as high up as $1 \,\rm{kpc}$ above the midplane at $R\sim 20 \, \rm{kpc}$, while in the solar neighbourhood the effect of the warp are much more modest ($0.1-0.2 \,\rm{kpc}$ variations in the mean height of the disc). The sun is situated at $(x,y)=(-8,0)$. Although the LMC is past its first pericenter, the interaction has not yet evolved to produce any ripples, just a warp. We will return to this in Section in 4 when discussing the impact of the Sagittarius dwarf on the disc.

\begin{figure}
\includegraphics[width=0.5\textwidth,trim=0mm 0mm 0mm 0mm,clip]{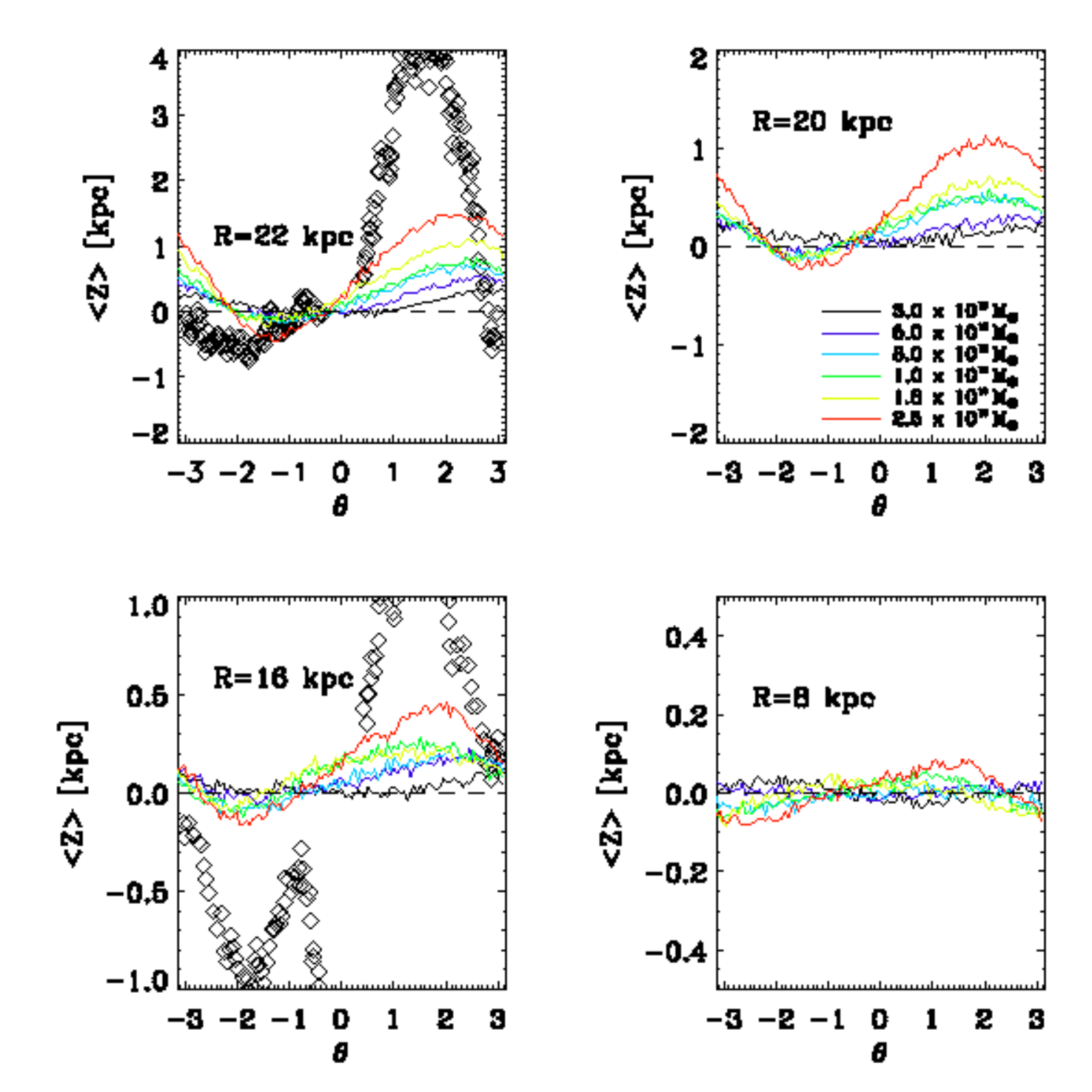}
\caption[]{Warp at the present day location of the LMC with respect to the Galactic center for all six models. Overplotted are the inferred HI data points from Levine et al. (2006) for the height of the gas disc above and below the midlplane. The LMC produces a warp that can be sufficiently characterised by three modes $m=0,1,2,3$ like the observed one, with qualitative features reminiscent of the observed HI warp (antisymmetry about the midplane with stronger amplitudes at $\theta=\pi/2$). Models of the LMC with infall mass $M_{vir}\geq2.5\times10^{10} \, \rm{M_{\odot}}$ are favoured by the data. }
\end{figure}

\subsubsection{Comparison to HI data}
We now compare in more detail features of our simulated discs to observations of the MW. This comparison will mostly revolve around the HI data of \cite{levine06} which has much larger spatial and angular coverage than stellar optical surveys (e.g. \cite{reyle09}). These authors infer the distance of gas elements by assuming elliptical orbits for the gas and a flat rotation curve at $220 \rm{km/s}$ everywhere. The vertical amplitude of the warp is calculated for each element through a mass-weighted mean height of the gas disc. As such, this is similar to calculating the mean height of the disc as we do in the simulations and both methods should yield the same answers - any discrepancy between the observations and the models will be related to either a real physical difference or systematics related to the derivation of the density grid and more crucially the inference of distances through the coordinate transformations performed by \cite{levine06} (see their eq. 1 and 2).

Figure 7 shows the mean height of the stellar particles as a function of azimuth for the six different LMC mass models considered. The diamonds represent the mean HI disc height from \citep{levine06}. Qualitatively, the discs perturbed by the more massive LMC models ($m_{LMC}\geq 10^{11}\,\rm{M_{\odot}}$) show similar features to the HI gas. A Fourier decompositions revealed that the disc can be sufficiently described by the superposition of three Fourier modes (m=0,1,2) of different phases. The $m=0$ introduces a constant shift about the midplane, the $m=1$ represent the first sinusoidal variation and the $m=2$ mode modulates the skewness of the principal $m=1$ sinusoid. Higher modes were much weaker and effectively absent between the simulations and observations.

Some differences are also visible in Figure 7. At $R\sim22\,\rm{kpc}$, there is an apparent phase offset between all models and the HI data of 0.4 radians ($\sim20$ degrees) which is much larger than the angular separation between the observed LMC's present-day position vector and those of the N-body models we considered. However, the exact phase of the disc response is highly sensitive to time: if we follow the response of the disc further in time by a few million years ($\sim0.05 \,\rm{Gyr}$) while still leaving the LMC within $2\sigma$ of its present day distance and speed to the galactic center as done in Figure 8, we see much better agreement with the phases both at $16 \,\rm{kpc}$ and $22\, \rm{kpc}$.

\begin{figure}
\includegraphics[width=0.5\textwidth,trim=0mm 0mm 0mm 0mm,clip]{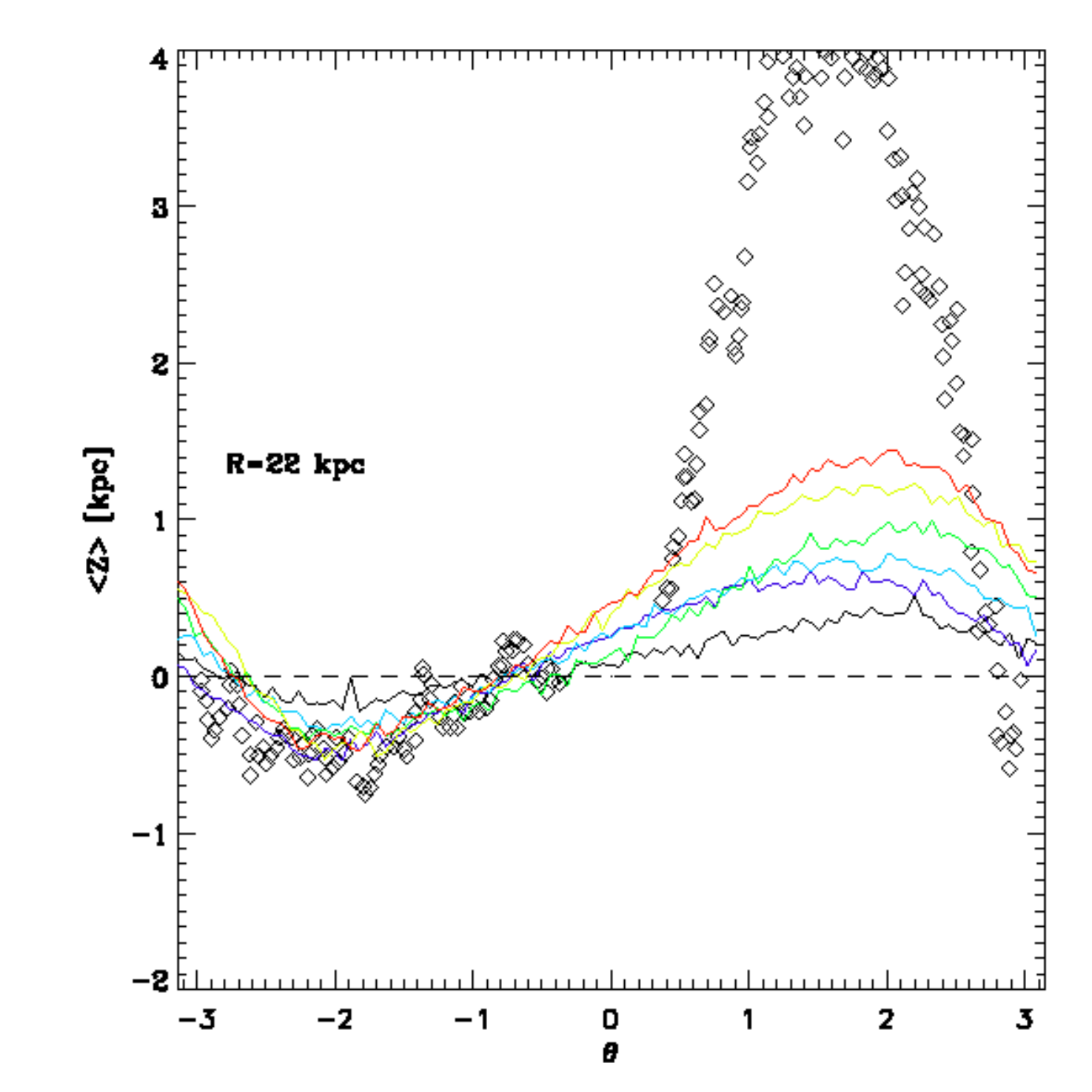}
\caption[]{Warp at a few timesteps later $\sim 0.05 \,\rm{Gyr}$. The phase-shift disappears as the disc had enough time to react, showing better agreement with the HI data in the phases. Discrepancies remain in the amplitude of the $m=1$ mode.}
\end{figure}

Nonetheless, the amplitudes of the observed and simulated warps also do not match the observations. While at $22 \,\rm{kpc}$, the models match the minimum crest, there is a deficit around the upper-crest. Even by considering the 96th percentile of particles, the models fall short by $1 \,\rm{kpc}$ and about $0.7 \,\rm{kpc}$ at $R=22 \,\rm{kpc}$ and $R=16 \,\rm{kpc}$ respectively. 

\subsubsection{Solar Neighbourhood constraints}
By considering masses of the LMC that are an order of magnitude larger than previously assumed (e.g. \citep{weinberg06}), it becomes important to check that the impact of the LMC on the disc does not affect constraints in our Solar Neighbourhood. We select disc particles within a circle of $1 \,\rm{kpc}$ radius in the plane at the assumed position of the Sun at $R=8 \,\rm{kpc}$. We then plot the mean velocity $\overline{v_{z}}$ as a function of height in the range $-1.6\le Z\le 1.6\, \rm{kpc}$ for our most massive model $m_{\rm{LMC}}=2.5\times10^{11} \rm{M_{\odot}}$. In the upper panel of Figure 9, we show that the fluctuations caused by the interaction with the massive LMC (red diamonds) vary between $ -2 \leq \overline{v_{z}} \leq 2 \,\rm{km/s}$ whereas the observed fluctuations in the Solar Neighbourhood taken from \cite{widrow12} are in the range $-10\leq \bar{v_{z}} \leq 10 \,\rm{km/s}$ (black triangles). Thus, such a model (or an even more massive one) is viable. 

Furthermore, the lower panel of Figure 9 shows the residuals in number density counts in the vertical direction. This is computed as $\Delta n(Z)= (n(Z)-n_{avg}(Z))/n_{avg}(N)$, where $n(Z)$ is the normalised number density measured within a cylinder of radius $1 \,\rm{kpc}$ around the Sun and $n_{avg}(Z)$ is the smooth normalised number density about a cylindrical annulus about $R=8 \\rm{kpc}$ of thickness $2 \,\rm{kpc}$ (so as to enclose the local Solar Neighbourhood cynlinder). Again, we see that the LMC cannot qualitatively produce a North-South asymmetry seen in \citep{widrow12}, where sinusoidal variations exist across the vertical direction in the disc. The blue diamonds represent points from a simulation of the Sgr dSph interacting with the MW, which we will discuss in section 4. While the mass of our favoured LMC model is high, we can conclude that it does not create disturbances as strong as those observed locally in the MW, making it allowed by the data and even leaving room for higher mass models.

\subsubsection{Summary}
So far, we note four important features confirmed by our N-body experiments. First, the observed phases of the response of the disk are well represented in the first passage scenario for the LMC. Secondly, the shape of the simulated stellar warp, $Z(R,\phi)$ at a given radius $R$, is not always symmetric about the midplane, like the observed warp traced by the HI. In the observations, the feature at $22 \rm{kpc}$ is anti-symmetrical due to a contribution from the m=0 mode. This is to be contrasted with the feature at $R=16 \,\rm{kpc}$ in the observations, which shows a weak m=0 contribution and a dominant m=1 signal. Third, we do not observe local disc disturbances that are unreasonably large compared to the data. Fourth, we observe a lack of agreement in the amplitudes of the vertical disturbances between our models and observations.

\begin{figure}
\includegraphics[width=0.5\textwidth,trim=0mm 0mm 0mm 0mm,clip]{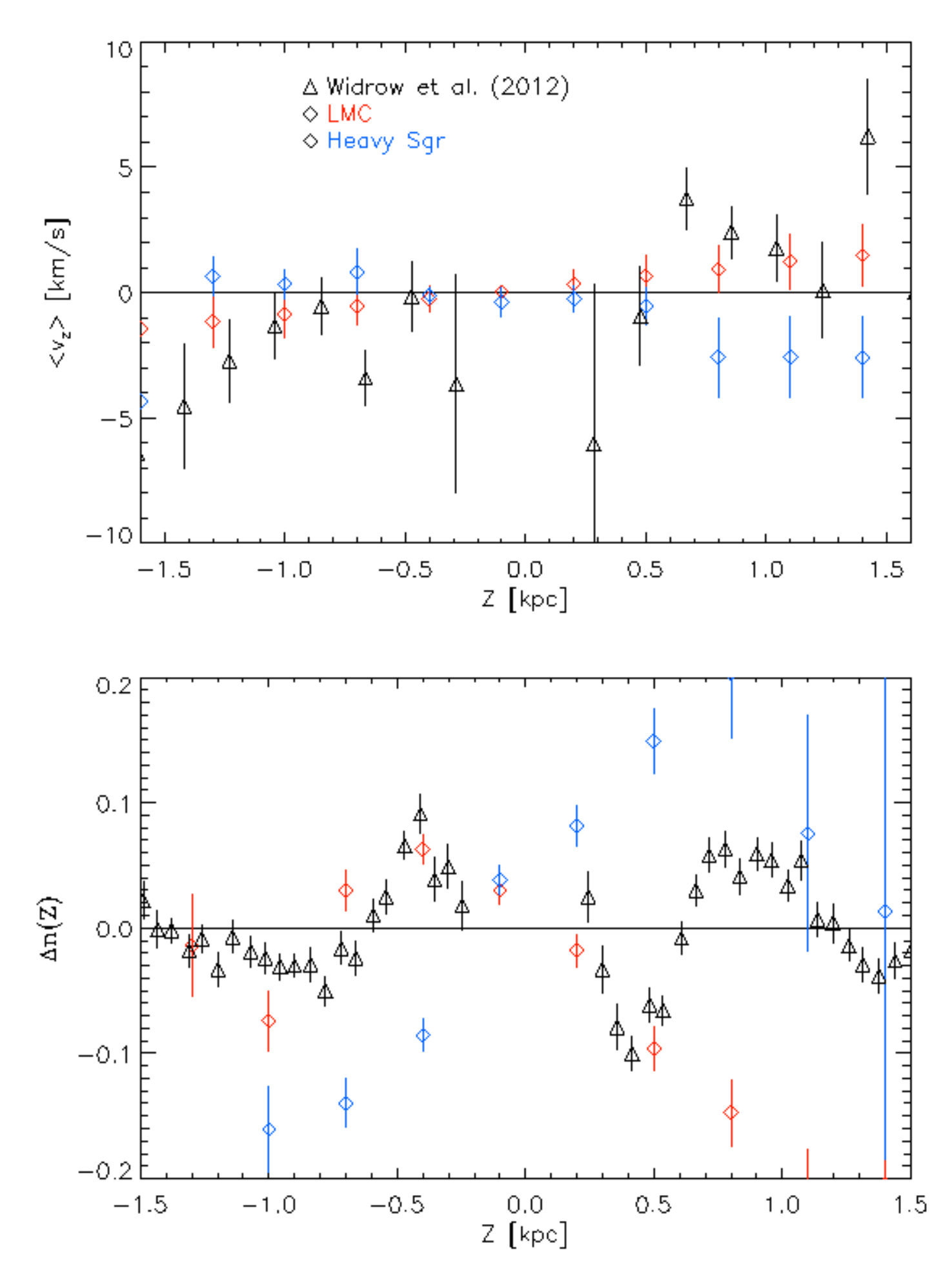}
\caption[]{{\it Top panel}: Mean vertical velocity about the midplane of the disc in the Solar Neighbourhood. The black triangles are observations from \cite{widrow12} while the red and blue diamonds correspond to the simulated signals for the LMC and Sgr respectively. The amplitudes of fluctuations in vertical bulk velocities of the heavy LMC model are systematically smaller than those observed in the observations. {\it Bottom panel:} Residual distribution $\Delta n(Z)$ as a function of height for a solar-like neighbourhood region. While Sgr is able to inflict fluctuations in the number density profile, this is not seen with our LMC model. The mismatch with the Widrow data was already noted in \citep{gomez13}.}
\end{figure}

\subsection{Hydrodynamical simulations}
A dissipational component such as cold gas in the disc can absorb some of the gravitational energy during the passage of the LMC and thus decrease the amount of energy transferred to the stellar component of the disc \citep{moster10}. We compare the results from the hydrodynamical run with our N-body run for a model where $m_{\rm{LMC}}=2.5\times10^{11}\rm{M_{\odot}}$ by looking at the amplitude of the warp at a radius of $20\, \rm{kpc}$ in Figure 10. The mean height of the gas is computed in the same way as we have done for the stellar disc in order to make a fair comparison between the two simulation runs. We note that the inclusion of gas results in a decrease of $0.1-0.3 \, \rm{kpc}$ in the amplitude of the mean heights of stars. This supports our suggestions above for even higher infall masses for the LMC. Furthermore, by including the HI we still recover the same features in the HI disc as we did for the stars in the dissipationless runs. However, given the much smaller mass contained in the HI disc, the gas warp is noisier due to poorer sampling of the underlying gas density.

\begin{figure}
\includegraphics[width=0.5\textwidth,trim=0mm 0mm 0mm 0mm,clip]{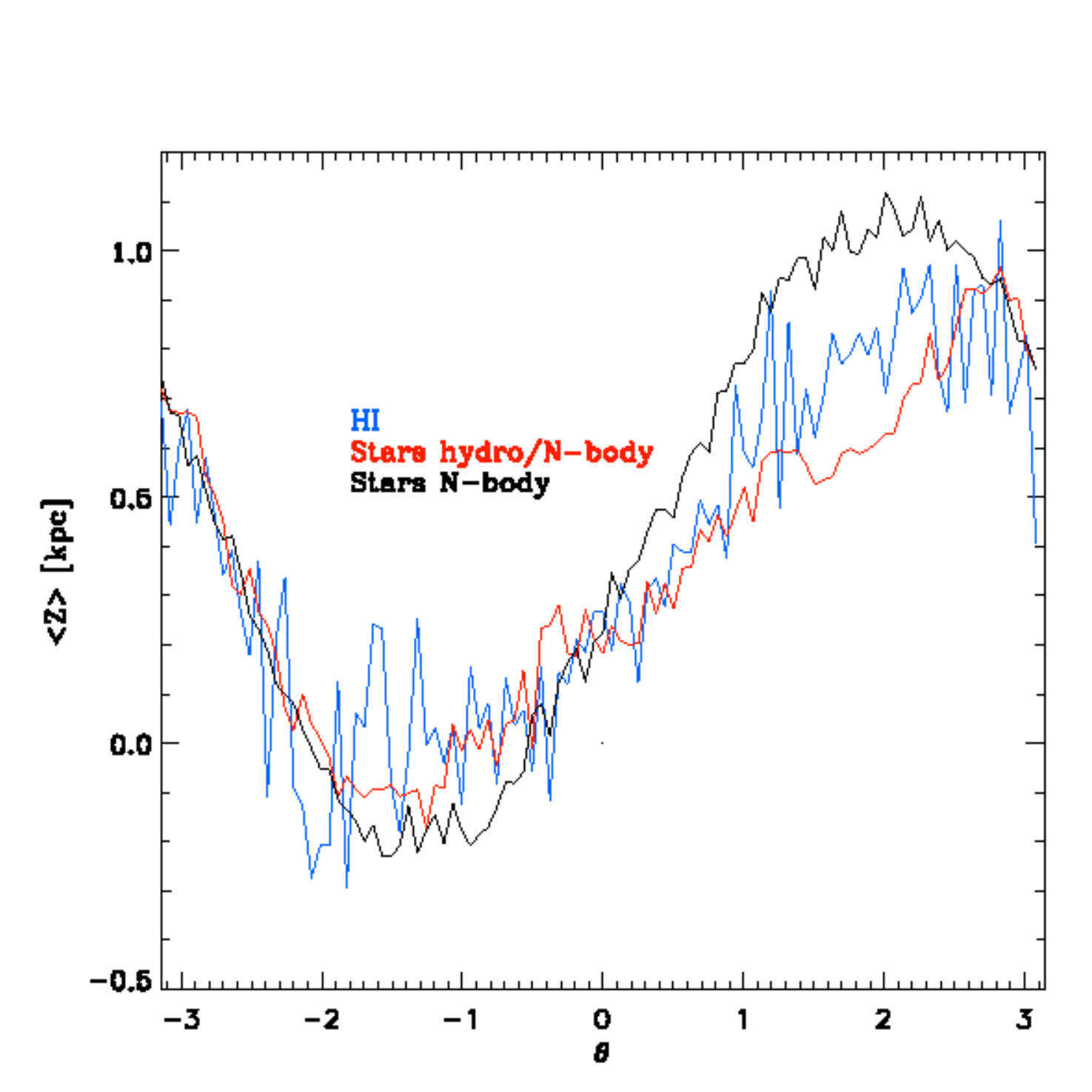}
\caption[]{Comparison between the hydrodynamics/N-body run and N-body run for the $M_{LMC}=2.5\times10^{11} \rm{M_{\odot}}$ model.}
\end{figure}

\subsection{The MW disc as a gravitational detector}

Assuming that the warp is due solely to the LMC and our mass model of the Galaxy (and the disc) are correct, the observed maximal disc height may be used to infer a mass for the LMC.  The idea of using the vertical structure of galactic discs to constrain the dark matter content of satellites is not new (see e.g. \citet{jiang99})\footnote{In a similar spirit \citep{chakrabarti09} proposed using the density profile of gas discs to constrain the existence of past interaction.}. The first observational application of ``Galactoseismology" was presented by \cite{widrow12}. In Figure 11, we observe a quasi-linear relationship between the mass of the LMC and the maximum height of the simulated disc (measured at $R=22 \,\rm{kpc}$) which could be used to put a lower limit on the mass of the LMC. Taking the relation in Figure 11 at face value, in order to explain the observed maximum mean height of the HI gas as observed in \citep{levine06}, we would infer an LMC infall mass of $M_{\rm{vir}}\sim5.0\times10^{11} \, \rm{M_{\odot}}$. This suggests that considering higher infall masses for the LMC is still possible and allowed by the data as shown in section 3.2. \cite{tsuchiya02} also observed that for a fixed LMC mass that the amplitude of warping increases with host halo mass. Thus, the relation we observe in our Figure 11 should define a manifold when taking the mass of the MW into account, leaving  a model with a more massive MW a viable solution. Another interpretation of the data could also be that misaligned infall of gas conspires together with the effect of the LMC's orbit on the disc to give rise to the observed distribution of HI gas. We will return to these issues in the discussion. 

\begin{figure}
\includegraphics[width=0.5\textwidth,trim=0mm 0mm 0mm 0mm,clip]{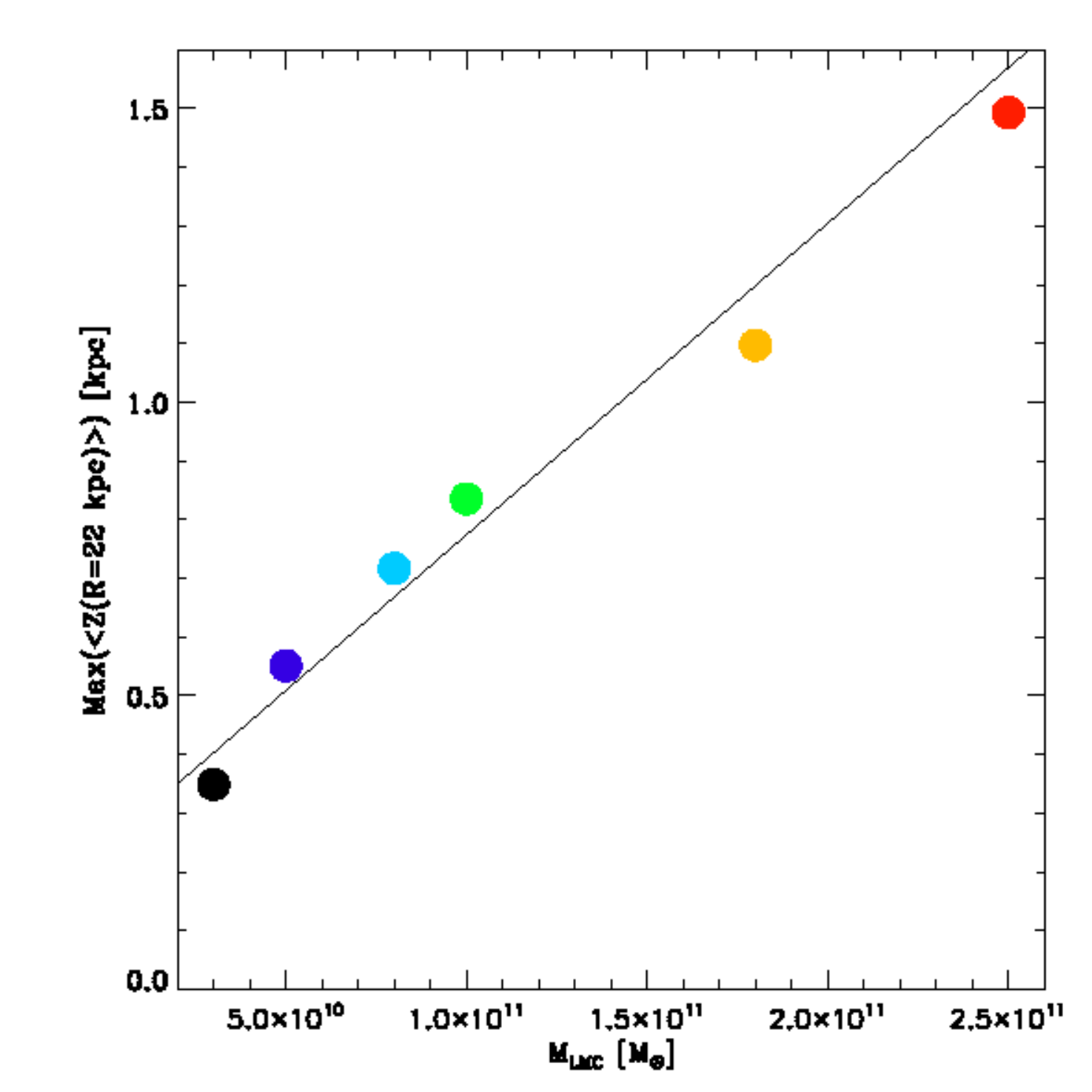}
\caption[]{Maximum heigh of the warp at a radial distance of $R=22 \,\rm{kpc}$ against the initial virial mass of the LMC for all six models. Because of the longer dynamical timescales in the outer disc, all these values remain distinct and strongly point towards a linear relation between initial mass of the progenitor and maximum disc height. This leaves the possibilities of using the present-day structure of the MW disc to place limits on the initial mass of the LMC.}
\end{figure}

\section{Results II: Comparison with the effect of Sgr}

In the previous sections, we found that the influence of massive LMCs on a first infall were consistent with the relatively unperturbed state of the local disc, while on larger scales the phase and sense of asymmetry in the warp could be explained by an interaction between the MW and the LMC.  However, we also noted that the amplitude and degree of asymmetry was not well matched and noticed that the scale of the interaction did not lead to the formation of rings/ripples such as the Monoceros Ring or the Triand clouds. One obvious reason is that the LMC alone is not the only architect of the the vertical structure of the MW disc.

In this section we compare and contrast the influence of another major player, the Sagittarius dSph which has sometimes been cited as an alternative potential candidate for exciting the warp of the MW \citep{bailin03}. Given that its orbit is perpendicular to that of the LMC, we might expect lines of node to be perpendicular to those produced by the LMC. However, such suppositions can be misleading given the orbit of Sgr and transient nature of warps (which winds itself up as time progresses). \cite{purcell11} studied the effect of the Sagittarius dSph on the disc of the MW. Both this and later studies showed that it could produce qualitatively asymmetries in the Solar Neighbourhood that are a the result of global bending model perturbing the entire disk \citep{gomez13}. \cite{price-whelan15} showed that ripples in outermost regions of the pre-existing disk could naturally account for Triand clouds. These simulations have caveats (e.g. see \cite{price-whelan15,gomez15b}) in particular, the initial conditions neglect the earlier phase of approach of Sagittarius in the MW, any earlier perturbations due to the dark matter halo wake produced by earlier and the first infall of Sagittarius are omitted by construction. Nevertheless, it is still instructive to compare the features left by Sagittarius on the disc to our models featuring the LMC.\\

Figure 12, compares the mean height of the disc as a function of azimuth for our most massive LMC run and our Heavy Sgr taken from \citep{price-whelan15}. This model assumes initial conditions similar to those used in \cite{purcell11}. Sagittarius is launched at $80 \,\rm{kpc}$ from the Galactic centre in the plane of the disc at $80 \,\rm{km/s}$ in towards the North Galactic Pole. In this simulation, the dark matter halos were represented by \cite{Navarro1996} profiles. The satellite mass has $M_{vir}=10^{11} \,\rm{M_{\odot}}$ and a scale length of $6.5 \,\rm{kpc}$. The MW model is slightly different in this model with a virial mass of $M_{vir}=10^{12} \rm{M_{\odot}}$ with scale radius $14.4\, \rm{kpc}$, a disc mass of $3.6\times10^{10}\,\rm{M_{\odot}}$ with exponential scale height of $2.84 \,\rm{kpc}$ and vertical scale height of $0.43 \,\rm{kpc}$. The bulge is modeled by a de-projected Sersic profile of index $n=1.28$, mass $9.52\times10^{9}\,\rm{M_{\odot}}$ and effective radius $0.56 \,\rm{kpc}$. The simulation reaches a present-day configuration after 2.3 Gyr.

Given that Sagittarius had the chance to perforate the disc numerous times, it is not a surprise that the signal in the warp should be far more complex than a simple superposition of three modes m=0,1,2 as seen in \cite{levine06}. However, it is interesting to note that the amplitudes in the variation of the mean disc height are comparable to those of our heaviest LMC run. Another interesting qualitative aspect is the similarity in amplitudes of the disc's mean height around the Solar Neighbourhood and at $R=16 \,\rm{kpc}$ (ignoring the smaller crest from Sagittarius due to a previous passage through the disc). Both Sgr and the LMC seem to produce shifts in $<Z>$ at the SN that are larger than observed (e.g. see the bottom panel of Fig. 9). If the individual responses are negatively (at least partially) interfering, the this could help to reconcile the models with the observations. This demonstrates that considering the combined contribution of both satellites could be of importance in order to model quantitatively the amplitudes seen in the MW warp and the more outer features of the disc (e.g. Monoceros, Triand I and II).

In Figure 13, we look at how the mean height about the disc mid-plane varies along a wedge of 16 degrees about the galactic center in the direction of the sun. Sagittarius does not warp the disc but makes it rippled (see also \citet{gomez13}) whereas the LMC imparts a less complex signal attributed to warps. The shape of the ripple is to first order sinusoidal and exponential as shown. Interestingly, the signal from our most massive LMC models follows the envelope of our heavy Sgr model. Because of the similarity in amplitudes of the perturbations imparted by both satellites, this clearly motivates future work considering the effect of both satellites and the coupling of signals (Laporte et al. in prep.).

\begin{figure}
\includegraphics[width=0.5\textwidth,trim=0mm 0mm 0mm 0mm,clip]{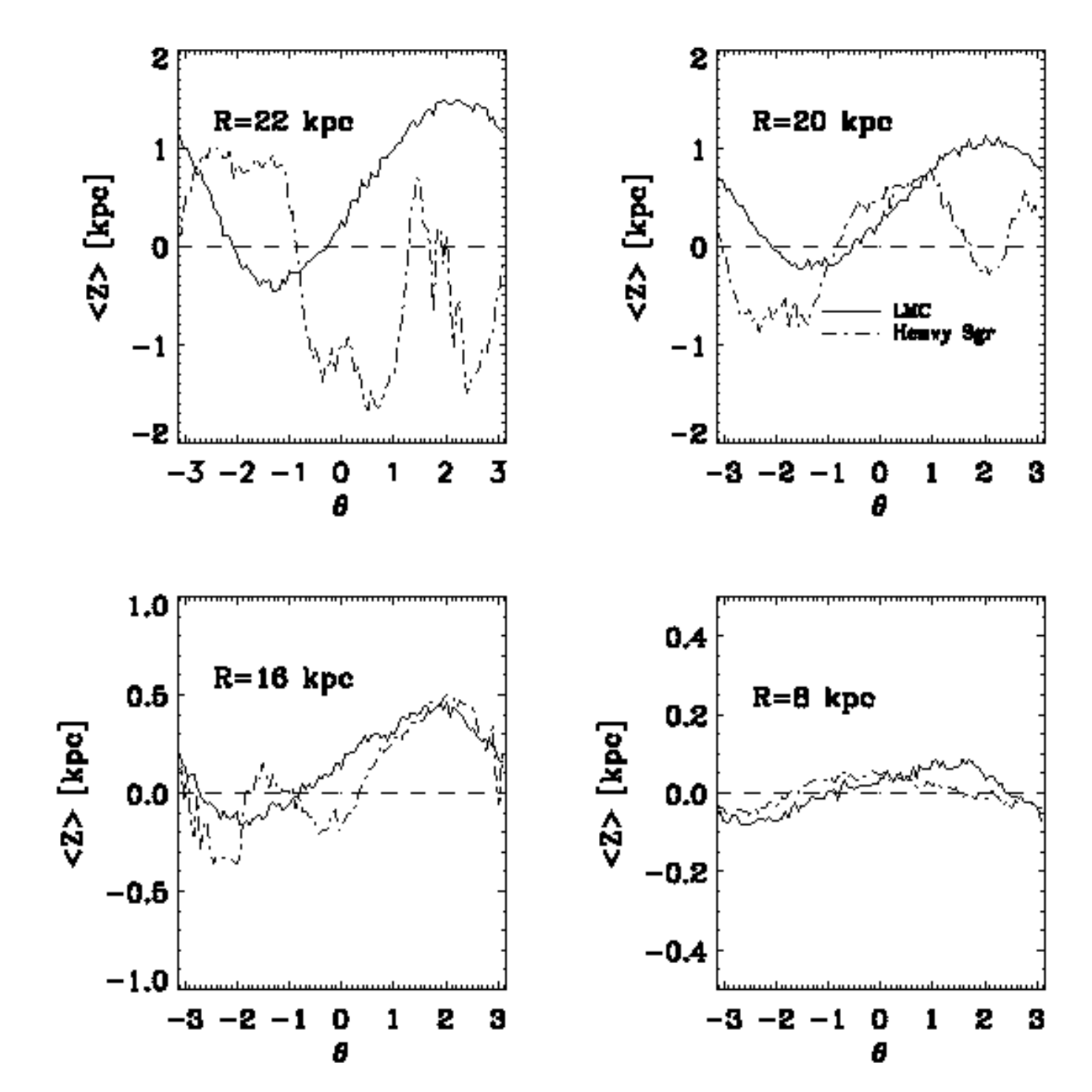}
\caption[]{Comparison between the Heavy Sgr run and N-body run for the $M_{LMC}=2.5\times10^{11} \rm{M_{\odot}}$ model. Sgr exihibits higher modes that are not seen in the observations and generally out of phase with the HI observations. At $R=16 \,\rm{kpc}$, the global m=1 mode Sgr is roughly of the same phase and amplitude as that for the LMC. Also around $R=8 \,\rm{kpc}$ the Sgr signal has a similar amplitude to that of the LMC.}
\end{figure}

\begin{figure}
\includegraphics[width=0.5\textwidth,trim=0mm 0mm 0mm 0mm,clip]{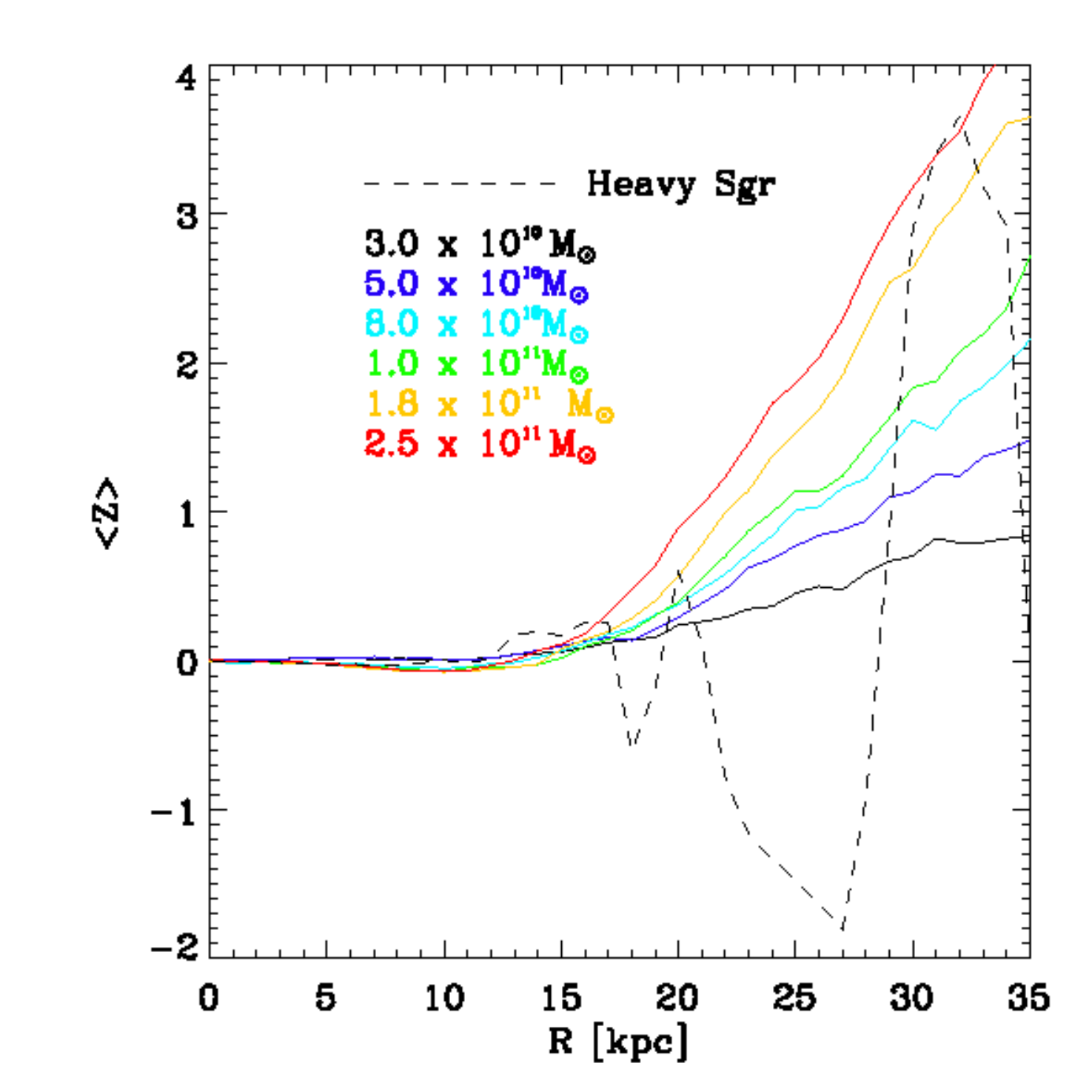}
\caption[]{Mean disc height as a function of galactocentric radius within a wedge of 16 degrees about the galactic center. There is an increase in mean height with distance which is also more pronounced the more massive the LMC is. The overplotted dashed line is the signal expected from a Sagittarius satellite with infall mass $10^{11} \rm{M_{\odot}}$ from a simulation using the same initial conditions as Purcell et al. (2011) used in Price-Whelan et al. (2015).}
\end{figure}

\section{Discussion}

\subsection{Mass of the LMC and its consequences for Galactic Archeology and stellar dynamics}
Our results favour an infall mass of $m_{LMC}\geq2.5\times10^{11} \,\rm{M_{\odot}}$, assuming the warp to be produced solely by the LMC. While historically unusual, this is not unreasonable given different lines of evidence. Using the Millennium-II simulation, \cite{boylan-kolchin10} tried to find Milky-Way analogues for which he used abundance matching relations of \cite{Guo2010} to infer an infall mass for the LMC assuming a stellar mass of $m_{*}=2.5\times10^{9} \rm{M_{\odot}}$. Considering the large scatter $\sim0.25 \,\rm{dex}$ in abundance matching relations reported in \citep{Moster2013} at those stellar masses, our preferred mass is well within the $+1\sigma$ limit ($m_{\rm{infall}}\sim3.4\times10^{11}\,\rm{M_{\odot}}$). Although abundance matching is not a predictive tool, we note that hierarchical galaxy formation models such as the Munich models \citep{henriques15} make predictions for the evolution of the luminosity function with redshift and how galaxies should populate dark matter halos finding similar relations to those required by abundance matching. On the dynamical side, \citep{penarrubia16} use the timing argument to deduce a mass of $m=2.5^{+0.09}_{-0.08}\times10^{11}\rm\,{M_{\odot}}$ for the LMC.

Indeed, looking at the amplitude of the warps produced for our different models, larger masses may still be conceivable. However, it is uncertain how much better a match to the observed HI data could be found (especially in the phases). Taking into account the random errors in the measured heights from \cite{levine06}, the discrepancy still persists. As a back of the envelope calculation, if we extrapolate the linear relation in Figure 8 to match the maximum of the observed warp amplitude at $R=22\, \rm{kpc}$ (taking into account the errors in Figure 8 of \citep{levine06}) we would deduce an infall mass of $M_{\rm{vir}}\sim5.0\times10^{11} \, \rm{M_{\odot}}$. At that same radius, a mass of $3.5\times10^{11}M_{\odot}$ would launch material to $Z\sim2.1 \, \rm{kpc}$. Whether such higher mass models would not violate vertical velocity constraints in the Solar Neighbourhood, remains to be tested. A larger mass of the LMC ($\geq 2.5\times 10^{11}\,\rm{M_{\odot}}$) would have important consequences for determining which of the newly discovered MW satellites \citep{koposov15,bechtol15, martin15} were once associated with it in the past \citep{deason15,jethwa16,sales16}. Moreover, a higher LMC mass would also imply a stronger tidal field as well as wake in the halo which could affect the motion of some globular clusters in the halo (Garavito et al. in prep.) and streams. \citep{gomez15}

\subsection{MW mass model}
The time for the infall of the LMC (and any other satellites) will be sensitive to the mass of the MW. In fact this quantity is quite uncertain by almost a factor of three \citep{xue08,Li08,watkins10,deason12,boylan-kolchin12, penarrubia14, penarrubia16}. Our simulations only considered a single mass model for the MW of $M_{200}=10^{12}\rm{M_{\odot}}$. A larger mass would allow, for example, the LMC to orbit in a less eccentric manner and be in the halo longer. As a result, this could affect the timescale of the response of the disc as well as the amplitude of the response. For example, \cite{tsuchiya02} observed that, for a fixed LMC mass and orbit, increasing the MW mass (he considered $M(<170\rm{kpc})=0.9\times10^{12}\, \rm{M_{\odot}}$) by a factor of two, could affect the maximum amplitude of the warp by a factor of two in the outer disc by roughly a factor of two. At face value, this could also be another way to reconcile the discrepancies we observed in our simulations. This will be investigated in future contributions. In order for the LMC to have made an orbit in in the halo, the MW would need to be in excess of $1.5\times 10^{12}\,\rm{M_{\odot}}$ (Patel, Besla et al. in prep) - such halo masses are not favored by the majority of recent mass estimates. 


\subsection{Possible errors in distance estimates in HI studies}
It is also important to note observational studies such as \citep{levine06, kalberla07} rely on assuming a flat rotation curve and using simple trigonometric relations to infer distances to HI gas. While this works reasonably well in the inner MW, it is not clear whether this is accurate at larger distances $R \geq16 \,\rm{kpc}$. Although this is beyond the scope of this paper,  exploring the systematics involved in such inference problems through mock observations of simulated data (such as our own) would be very informative for both observers/modelers and theorists, to understand the shape of the HI gas warp in the MW. Are the features at $\sim 22 \,\rm{kpc}$ real? If so, is it a signature of the effect of the LMC on the disc or a combination between the LMC and recent cold gas accretion onto the MW?
\subsection{Alternative explanations}
Other proposed mechanisms to create the warp in the MW have been proposed. Indeed, \cite{dubinski09} considered simulations of the Millky Way evolving in the presence of tidal torques due to a misalignment between the disc and dark halo, in the form of a quadrupole. Interestingly, they showed that for reasonable parameters taken from cosmological simulations, they could vary the mean height of the outer disc to larger values than reported here. However, the level of misalignment between the dark halo and the disc of the MW is completely unconstrained by current observations. Furthermore, a recent study based on a suite of high resolution fully cosmological simulations showed that such misalignments are very unlikely to be the main drivers behind large vertical perturbations in galactic discs \citep{gomez16}.

Studying the impact of both the triaxiality (in particular the impact of a misalignment between the halo and disc) subject to the infall of the LMC would be another interesting possible avenue to consider in order to explain the shape and amplitude of the observed HI warp. \cite{gomez15b} present a case study where they find that such misalignments present in cosmological hydrodynamical simulations of disc formation are weaker than wakes produced by fly-by encounters.  Misaligned cold gas accretion may also be another mechanism to explain the large mean heights seen in the HI gas \citep{roskar10, gomez16}. \cite{chakrabarti09} explored the tidal imprints on the gas surface density of a passing hypothesized dark subhalo with mass ratio 1:100 and $5 \,\rm{kpc}$ pericentric radius (now at $90\,\rm{kpc}$) to explain the existence high order modes present in the measured surface density maps \citep{levine06b}. However, the authors did not report on the vertical structure of their HI disc.

\subsection{The case for a menage a trois}
It should be noted that we just took a simple mass model of the MW and that our models were not tailored to model the warp a priori. The pronounced vertical perturbation was a result of the tidal interaction of the LMC on the disc and the response of the MW dark matter halo to the wake produced by the LMC.  Future modeling of the LMC and MW systems will need to further address the detailed shape of the vertical perturbations to the disc (in phase and amplitude) in order to trace the history of the MW and infer the mass of the LMC at infall. However, this cannot be the complete picture. The vertical structure of the stellar disc is not only warped, but also shows signs propagating rings/ripples \citep{newberg03,xu15,price-whelan15, morganson16} which can be qualitatively reproduced when considering the effect of the Sgr dSph. While in our current study, we have not modeled the combined effect of Sgr and the LMC. We demonstrated that both separate models (LMC+MW and Sgr+MW) can individually produce features of similar to amplitude that may, in some regions, conspire in phase. To our knowledge, the combination of effect induced by the LMC and Sgr remains the most compelling culprit for the vertical perturbation of the MW stellar disc and we have presented models that show much promise for future experimentations. 

We should also note that most N-body experiments consider initially stable discs (as was the case for our own study). This is certainly not the case for the MW (as Sgr has been orbiting the MW for a longer period of time than the LMC) or expected given how structure forms in $\Lambda$CDM. Thus, the response of an initially perturbed disc (through a merger or other processes) ought to be investigated. Given that the masses of the LMC and Sagittarius may be revised towards higher values than previously considered\footnote{\citep{jiang99} was the first study to propose a mass for Sgr as high as $10^{11} \,\rm{M_{\odot}}$.} (.e.g \citet{besla07,gibbons16}), it is likely that the N-body method will become the method of choice to infer the past history of the Galaxy in its full complexity.  As our knowledge of galactic structure (disc kinematics, shape of the DM halo as a function of radius) will increase in the upcoming years towards a fiducial background model for the MW, we should hopefully be able to reconstruct the past history and interactions of the MW with its most massive satellites.

\section{Summary \& Conclusion}


Owing to the revised measurements of the LMC's proper motion \citep{kallivayalil06,kallivayalil13}, we studied the impact of the LMC of the galactic disc during a first infall passage. We have explored six different models for the LMC increasing in mass from $3\times10^{10}{\rm{M_{\odot}}}$ to $2.5\times10^{11} {\rm M_{\odot}}$. In agreement with HI maps of \cite{levine06}, the warp of the MW is fully characterised by the super-imposition of three modes (m=0,1,2). This was also shown in an earlier study by \cite{weinberg06} who used a matrix method \citep{weinberg98} to follow the evolution of the MW disc's distribution function to first order. We find that the maximum amplitude of the warp scales linearly with the perturber's mass. Our models correctly predict the phase of the MW warp, though we show that this feature is sensitive to the integrated time chosen. Although we do observe a discrepancy in the amplitude of our warps, our models favour a high infall mass for the LMC $m_{LMC}\sim2.5\times10^{10}\,\rm{M_{\odot}}$ or higher.

We show that a massive LMC does not violate spatial and kinematical constraints in the vertical direction in the Solar Neighbourhood leaving models with $m_{LMC}\geq 2.5e11\,\rm{M_\odot}$ allowed by the data. Comparing the responses of the disc to Sagittarius and the LMC, we show that the LMC is the most likely suspect to warp the MW as Sagittarius introduces higher order modes detected in the stars but not seen in the HI. However, our study suggest that the coupling between the modes excited by these two satellites could be important to reconcile differences between our models and observed features. Examples of this are the amplitude of the HI warp at $R\sim 16\,\rm{kpc}$ and the oscillatory behaviour observed in the stellar component of the disc. This is beyond the scope of the present paper and will be presented in a separate contribution focussing on the coupling of different modes in the response of the MW disc (Laporte et al. in prep) and testing whether perturbations from multiple satellites can be disentangled .

\section*{Acknowledgments}

We thank Volker Springel for giving us access to the {\sc gadget-3} code. CFPL is supported by a Junior Fellow award from the Simons Foundation and thanks Jacqueline van Gorkom, David Hogg, Robyn Sanderson and Adrian Price-Whelan for valuable discussions. This work used the Extreme Science and Engineering Discovery Environment (XSEDE), which is supported by National Science Foundation grant number OCI-1053575. We also acknowledge use of computing facilities at the Rechenzentrum Garching (RZG) and the Max Planck Institute for Astrophysics (MPA). KVJ's contributions were supported by NSF grant AST-1312196. N.G-C is supported by the McCarthy-Stoeger scholarship from the Vatican Observatory.

\bibliographystyle{mn2e}
\bibliography{master2.bib}{}
\label{lastpage}
\end{document}